\documentclass[final,12pt,3p,times]{elsarticle}

\usepackage{amsmath,amssymb,amsfonts}
\usepackage{mathrsfs}
\usepackage{chemformula}
\usepackage{color}
\usepackage[normalem]{ulem}
\usepackage{stmaryrd}
\usepackage{diagbox}

\biboptions{sort&compress}

\newcommand{\beq}{\begin{equation}}
\newcommand{\eeq}{\end{equation}}

\newcommand{\ep}{\varepsilon}
\newcommand{\Pra}{\textit{Pr}}

\newcommand{\mc}[1]{\mathcal{#1}}

\begin{document}

\begin{frontmatter}

\title{A simplified model for coupling Darrieus--Landau and diffusive-thermal instabilities}

\author{Prabakaran Rajamanickam}
\address{Department of Mathematics, University of Manchester, Manchester M13 9PL, United Kingdom\\
Email:prabakaran.rajamanickam@manchester.ac.uk}

\begin{abstract}
A simplified phenomenological model is proposed to couple the long-wave Darrieus--Landau (DL) instability and the short-wave diffusive-thermal (DT) instability in premixed flames. By identifying a cubic coupling term in the linear dispersion relation, representing the leading-order interaction between hydrodynamic expansion and diffusive transport, this framework moves beyond the traditional treatment of these instabilities in isolation. Two distinct asymptotic regimes are identified: the first recovers the classical Michelson--Sivashinsky equation for order-unity positive Markstein numbers $\mc M>0$, the second reveals a distinguished DL-DT crossover regime where both instabilities participate at equal order. In this crossover limit, where the Markstein number is small ($\mc M \sim \sqrt{\ep}$ with $\ep$ measuring thermal expansion), a generalized evolution equation is derived featuring a nonlocal stabilising term controlled by the hydro-diffusive number $\mathcal{N} = \mc A/\delta_L^2$, where $\mc A$ is the hydro-diffusive area---the characteristic area over which hydrodynamic and diffusive transport processes interact. This term remains active even when Markstein stabilisation vanishes. Numerical solutions in sufficiently large domains based on our model reveal a distinctive chaotic regime in which the characteristic DL cusp structures are in persistent competition with small-scale wrinkles. This minimal unified framework thus captures the essential coupled dynamics governing flame front instability and provides a tractable explanation for the fine-scale cellular structures and accelerated growth rates observed, without recourse to the full complexity of the complete conservation equations.
\end{abstract}


\end{frontmatter}

\section{Introduction}

The theory of instability in premixed flames has, for the most part, evolved along two parallel and only partially intersecting lines of inquiry. The first of these, originating in the classical analyses of Darrieus~\cite{darrieus1938propagation} and Landau~\cite{landau1944slow}, establishes that a plane flame, by virtue of the density change induced through thermal expansion, is intrinsically susceptible to a hydrodynamic instability of long wavelength.  The second instability, foreseen by Zeldovich~\cite{zeldovich1944theory} in the same era, arises from the disparity between thermal and molecular diffusion, and bears a formal resemblance to the Turing instability. The rigorous mathematical development of the Darrieus--Landau (DL) and diffusive-thermal (DT) instabilities saw a significant revival in the late 1970s and early 1980s through the pioneering contributions of Sivashinsky, Clavin, Matalon, Joulin, and their collaborators. Even today, these instabilities are largely treated separately rather than within a unified framework. While the DL instability persists in premixed flames under general conditions, the DT instability arises only for sub-unity Lewis numbers, due to preferential transport of heat relative to species diffusion~\cite{williams2018combustion,clavin2016combustion}.

The Matalon--Matkowsky--Clavin--Joulin (MMCJ) theory ~\cite{clavin1982effects,matalon1982flames,clavin1983premixed,pelce1988influence} has provided a definitive account of the DL instability, revealing how the inner structure of the flame imparts a stabilising influence at short wavelengths, of the order of the flame thickness.  In a parallel development, a quantitative theory of the diffusive-thermal instability was formulated by Sivashinsky, Clavin, and Joulin~\cite{clavin1985effect,sivashinsky1977diffusional}, but in this treatment the full hydrodynamic effects were suppressed, an approximation (i.e., thermo-diffusive approximation) that eliminates the DL mechanism. A comprehensive treatment in which both mechanisms emerge naturally from the full set of conservation equations remains, even now, an undertaking of considerable difficulty. It is of some historical interest that Sivashinsky, in his seminal contribution of 1977~\cite{sivashinsky1988nonlinear} (see also the review of Matalon~\cite{matalon2007intrinsic}), had already envisaged a regime in which the two instabilities might be coupled, though of a character distinct from that considered here.

In addition to theoretical investigations, a substantial body of numerical and experimental work has been devoted to the study of these instabilities, both singly and in combination; see, for example,~\cite{jackson1984effect,denet1992numerical,denet1995numerical}.
 In light of the complexity of a complete theory, it is desirable to construct a simplified phenomenological model that couples these instabilities. Such a model can provide insight into the interaction between DL and DT modes without requiring the full, detailed analysis. The purpose of the present note is to advance such a model. Crucially, while previous models often treat these instabilities as additive, we identify a 'hydro-diffusive' coupling that becomes dominant near the stability threshold. The objective of this short note is precisely to present such a minimal coupled model, introducing a new physical parameter, the hydro-diffusive area, to characterise this interaction.

\section{A modified dispersion relation near the instability onset (weak heat release)}

For convenience, we use the laminar flame thickness $\delta_L$ as the characteristic length scale, and $\delta_L / S_L$ as the time scale, where $S_L$ is the laminar flame speed. Near the onset of the DL instability, the dispersion relation can be written as
\begin{equation}
    \sigma_{DL} = \ep |k| -\mc M k^2 \qquad \text{with} \qquad \ep=\frac{r-1}{2}>0, \, \mc M>0, \label{DL}
\end{equation}
while for the DT instability,
\begin{equation}
    \sigma_{DT} = -\mc M k^2 - 4 k^4 \qquad \text{with} \qquad \mc M<0. \label{DT}
\end{equation}
Here, $r>1$ is the unburnt-to-burnt gas density ratio, and $\mc M$ is the Markstein number, which depends on the Lewis number. When $\mathcal{M}<0$, the DL relation breaks down, indicating the emergence of the DT instability.

A simple combination of these two linear relations might suggest~\cite{sivashinsky1988nonlinear,matalon2007intrinsic}
\begin{equation}
    \sigma = \ep |k| -\mc M k^2 - 4 k^4,
\end{equation}
but this form neglects the leading-order coupling between the long-wave DL and short-wave DT instabilities. A more appropriate  form (after noting $\sigma(k)$ must be an even function of $k$) may then appear to be $\sigma = \ep |k| -\mc M k^2 - \mc N |k|^3 - 4 k^4$. The new term $\mc N |k|^3$ represents the next-order stabilisation in~\eqref{DL} when $\mc M<0$, with $\mc N>0$. When the cubic term alone provides sufficient short-wavelength saturation, the quartic term is unnecessary. Therefore, the final form of the linear dispersion relation can be written as
\begin{equation}
    \sigma = \ep |k| -\mc M k^2 - \mc N |k|^3, \qquad \text{where} \qquad  \ep> 0, \qquad \mc M \in \mathbb{R}, \qquad \mc N>0.\label{disp}
\end{equation}
The term $-\mc N|k|^3$ can be interpreted as both the intrinsic hydrodynamic saturation of the DL branch and the leading-order coupling between long-wave DL modes and short-wave DT modes; it is the lowest-order nonlocal correction consistent with symmetry. Thus, in addition to the classical Markstein number $\mc M$, which characterises the linear response of the flame speed to curvature, we introduce a new parameter, $\mc N\sim O(1)$, referred to as the \textit{hydro-diffusive number}. While the Markstein number accounts for purely diffusive-thermal effects and has dimensions of length, $\mc N$ has dimensions of area (length squared), representing the higher-order interaction between the flow field and the internal flame structure. Thus we may write $\mc N = \mc A/\delta_L^2$, with $\mc A$ representing a (dimensional) \textit{hydro-diffusive area}, the characteristic area over which hydrodynamic and diffusive transport processes interact. It is important to emphasize that the cubic term $-\mc N|k|^3$ represents a higher-order correction to the DL dispersion relation that becomes negligible when $\mc M\sim O(1)$, but emerges as the dominant stabilising mechanism in the limit $\mc M\to 0^\pm$, where the classical quadratic stabilisation vanishes.

The dispersion relation~\eqref{disp} implies that the range of unstable wavenumbers are given by $|k|\in(0,k_c)$ and the  maximum growth rate $\sigma_m$ occurs at the wavenumber $k_m$, where 
\begin{equation}
    k_c = \frac{\sqrt{\mc M^2+4\ep\mc N}-\mc M}{2\mc N}, \qquad
    k_m = \frac{\sqrt{\mc M^2+3\ep\mc N}-\mc M}{3\mc N}. \label{wave}
\end{equation}

\section{Weakly nonlinear evolution: Two regimes}

The instability onset threshold is dictated by the limit $\ep\to 0^+$. Depending on how $\mc M$ scales with $\ep$, we can consider two distinct regimes where a weakly nonlinear description can be provided. Our objective is here to write down the governing equations for the flame amplitude $f(x,t)$, i.e., departure of flame excursion from the planar state, following the  approach in~\cite{rajamanickam2024tricritical,daou2024diffusive}. 

\subsection{The classical Michelson--Sivashinsky equation: $\mc M \gg \sqrt {\ep\mc N}$}

When $\mc M \gg \sqrt {\ep\mc N}$, the cubic term in~\eqref{disp} is negligible and unnecessary so that
\begin{equation}
    \sigma = \ep |k| -\mc M k^2, \qquad \mc M\sim O(1).
\end{equation}
Then scaling laws are given by
\begin{equation}
    k\sim \ep, \qquad \sigma \sim \ep^2, \qquad f \sim 1,
\end{equation}
in which the scaling law for $f\sim \sigma/k^2$ corresponds to the weakly nonlinear regime, i.e., $f_t \sim \tfrac{1}{2}f_x^2$. In this case, one covers the classical Michelson--Sivashinsky equation~\cite{clavin2016combustion}
\begin{equation}
f_t + \tfrac{1}{2}f_x^2 = \mc M f_{xx} +\ep \mathcal{H}(f_x), \label{MS}
\end{equation}
where $\mathcal{H}$ denotes the Hilbert transform, defined in Fourier space as $\mathcal{F}\{\mathcal{H}(f)\}(k) = -i\operatorname{sgn}(k)\hat{f}(k)$, so that $\mathcal{F}\{\mathcal{H}(f_x)\}(k) = |k|\hat{f}(k)$. 

\subsection{The DL-DT crossover regime: $\mc M \sim \sqrt{\ep\mc N}$ with $\mc N\sim O(1)$}

The DL-DT crossover regime is a distinguished regime in which both the Darrieus--Landau instability and diffusive-thermal instability participate at equal order.. This regime also includes the special case where $\mc M=0$. Let us introduce the order unity parameter
\begin{equation}
    \mu = \frac{\mc M}{\sqrt{\ep}} \sim O(1),
\end{equation}
so that the dispersion relation~\eqref{disp} can be written as
\begin{equation}
    \sigma = \ep |k| -\sqrt{\ep} \mu k^2 - \mc N |k|^3.
\end{equation}
From this, the scaling laws follow as
\begin{equation}
    k\sim \sqrt\ep, \qquad \sigma \sim \ep\sqrt\ep, \qquad f\sim \sqrt\ep.
\end{equation}
The amplitude equation for the crossover regime is given by
\begin{equation}
f_t + \frac{1}{2}f_x^2 = \mc M f_{xx} + \ep\mathcal{H}(f_x) + \mc N \mathcal{H}(f_{xxx}), \label{DLDT}
\end{equation}
where we used $\mathcal{F}\{\mathcal{H}(f_{xxx})\}(k) = -|k|^3\hat f(k)$. Note that in this regime $\mc M = O(\sqrt\ep)$ is small, consistent with the scaling.

Several important features distinguish this regime from the classical MS case:
\begin{itemize}
    \item \textbf{Wavenumber}: $k \sim \sqrt\ep$, which is \emph{larger} than in the classical MS regime ($k \sim \ep$). This indicates that the cellular patterns are finer, with shorter characteristic wavelengths. The DT mechanism, even when weak, selectively amplifies shorter waves, leading to a more refined cellular structure compared to the pure hydrodynamic instability.
    \item \textbf{Growth rate}: $\sigma \sim \ep^{3/2}$, which is \emph{larger} than in the classical MS regime ($\sigma \sim \ep^2$). The combined influence of both instabilities produces faster growth than either mechanism in isolation. The coupling between DL and DT modes accelerates the instability development.
    \item \textbf{Amplitude}: $f \sim \sqrt\ep$, which is \emph{smaller} than in the classical MS regime ($f \sim 1$). The flame deformation is much smaller than the flame thickness, making this a more perturbative regime. The stronger short-wave stabilisation from the $\mc N|k|^3$ term limits the flame excursion more severely than in the pure DL case.
    \item \textbf{Critical wavenumbers:} The two critical wavenumbers in~\eqref{wave} now takes the form $k_c = \sqrt{\ep}(\sqrt{\mu^2+4\mc N}-\mu)/(2\mc N)$  and $k_m = \sqrt{\ep}(\sqrt{\mu^2+3\mc N}-\mu)/(3\mc N)$. As $\mu \to +\infty$, we approach the nearly planar waves $\{k_c,k_m\} \to  0$ and as $\mu \to -\infty$, we approach the finite wavelength cells $\{k_c,k_m\} \to  +\infty$. 
\end{itemize}
The crossover regime thus exhibits a unique combination of finer cellular structures, faster growth rates, and smaller amplitudes, a direct consequence of the coupled DL-DT dynamics captured by the distinguished limit $\mc M \sim \sqrt{\ep \mc N}$.

\subsection{Heuristic approach to the crossover regime, $\mc M \sim \sqrt{\ep\mc N}$, when $\mc N$ itself is small}

The crossover regime when $\mc N$ is a small number lies in the deep asymptotic limit $\mc M\to 0$.  One may attempt a heuristic phenomenological model that bridges the Michelson--Sivashinsky equation (describing the DL-dominated regime) and the Kuramoto--Sivashinsky equation (describing the DT-dominated regime). The idea is to retain all terms that appear in the linear dispersion relation, including the ultimate short-wavelength cutoff provided by the flame thickness, which would enter as a $-4k^4$ term in Fourier space. A minimal phenomenological extension that captures the combined effects is
\begin{equation}
f_t + \frac{1}{2}f_x^2 = \mc M f_{xx} + \ep \mathcal{H}(f_x) + \mc N\mathcal{H}(f_{xxx}) - 4 f_{xxxx}. \label{pheno}
\end{equation}
This equation, which pertains to the linear dispersion relation $\sigma=\ep |k|-\mc M k^2 - \mc N|k|^3-4k^4$, implies the scaling laws
\begin{equation}
    k\sim \ep^{1/3}, \qquad \sigma \sim \ep^{4/3}, \qquad f\sim \ep^{2/3}, \qquad \mc M\sim \ep^{2/3}, \qquad \mc N \sim \ep^{1/3}.
\end{equation}
That is to say, both $\mc M$ and $\mc N$ must be small numbers, although $\mc N \gg \mc M$. Such a scenario, in which both coefficients vanish simultaneously with specific scaling exponents, is a special (codimension-two) coincidence in parameter space. Generically, as the DT threshold is approached ($\mc M \to 0$), the hydro-diffusive number $\mc N$ is expected to remain order unity unless a separate fine-tuning occurs. Therefore, the deep asymptotic limit described here is unlikely to be realised in practice; the generic crossover regime is that discussed in Section~3.2.

\begin{figure}[h!]
\centering
\includegraphics[scale=0.5]{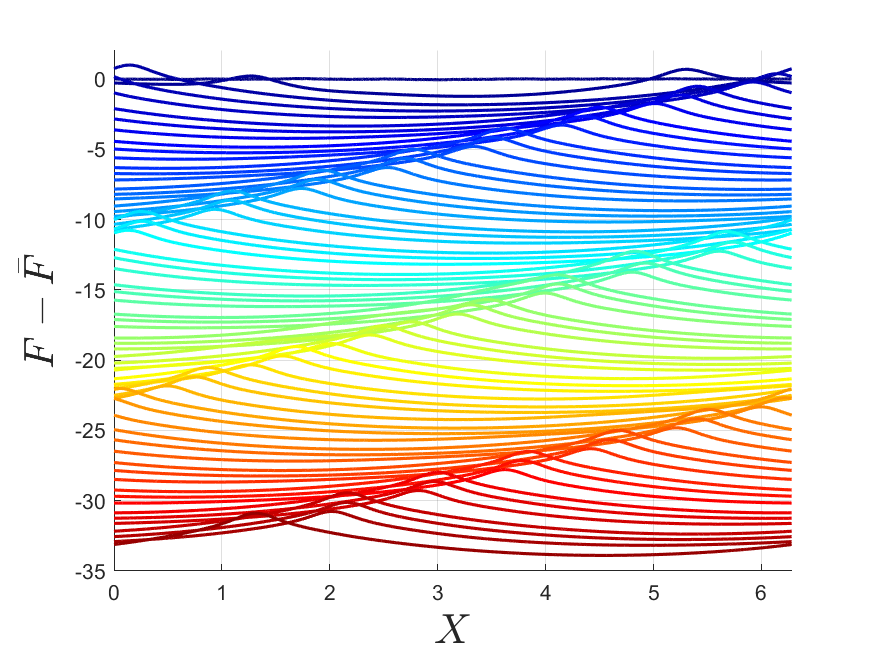} \hspace{1cm}
\includegraphics[scale=0.5]{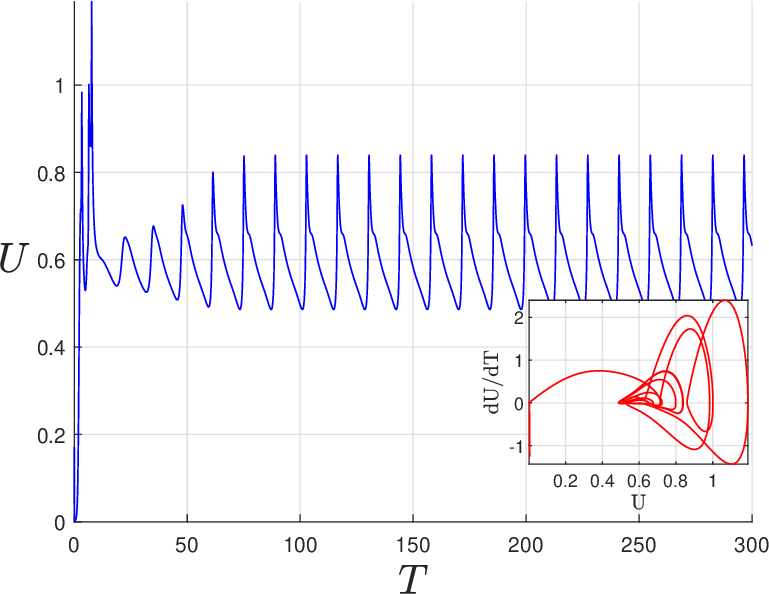} 
\includegraphics[scale=0.5]{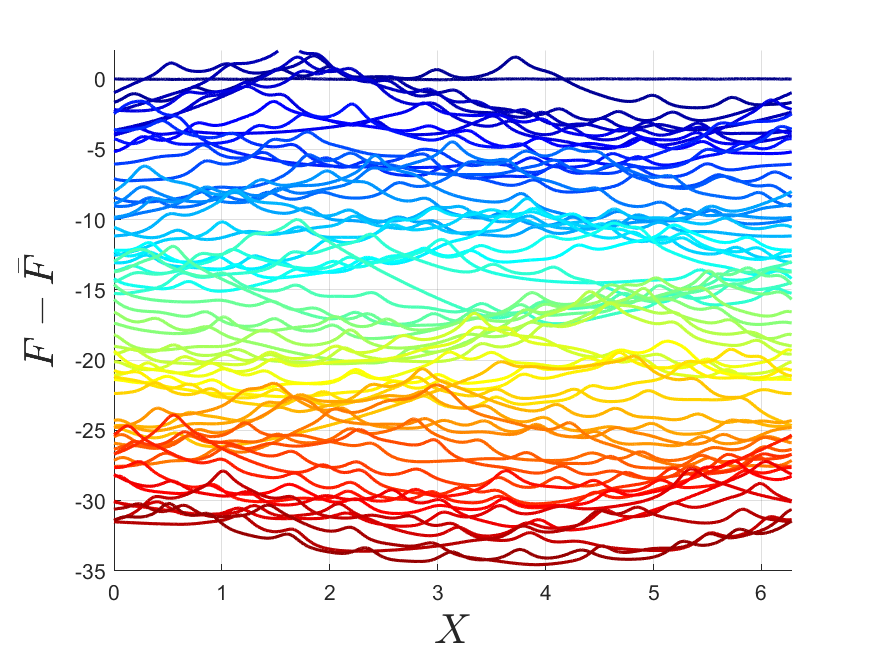} \hspace{1cm}
\includegraphics[scale=0.5]{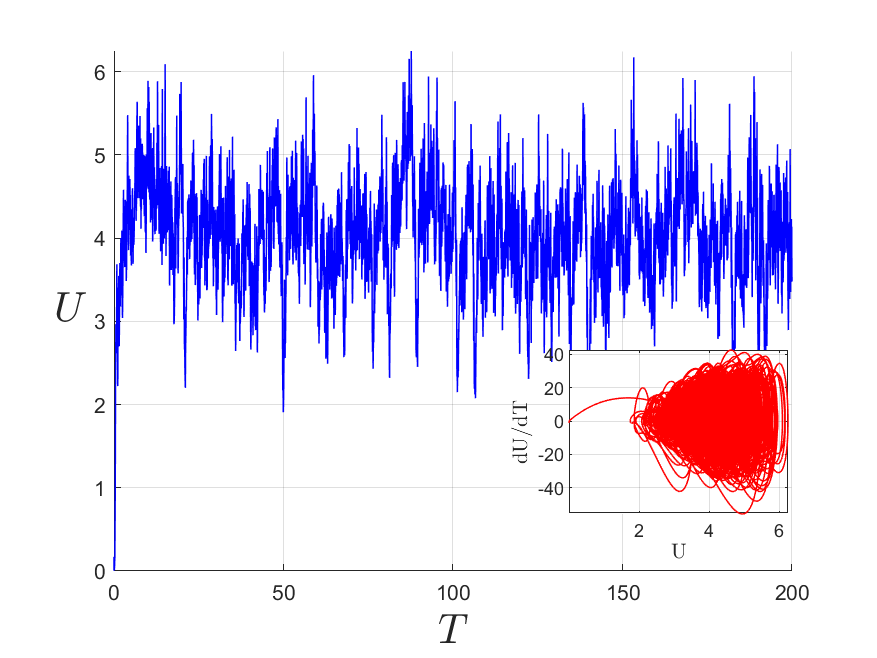} 
\caption{Numerical solution in a relatively large domain ($\nu=0.01$) with $\mu=+1$ (top row) and $\mu=-1$ (bottom row). Plotted are time snapshots of the flame shape and the trend of the propagation speed $U$; the small inset corresponds to the phase portrait, i.e., $U$ versus $dU/dT$. }
\label{fig:largedomain}
\end{figure}

\begin{figure}[h!]
\centering
\includegraphics[scale=0.5]{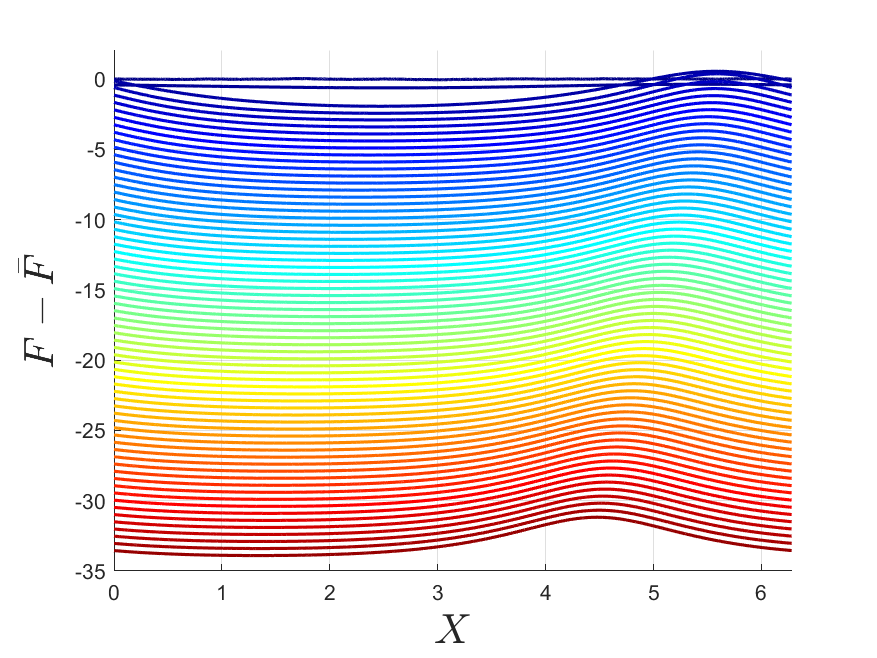} \hspace{1cm}
\includegraphics[scale=0.5]{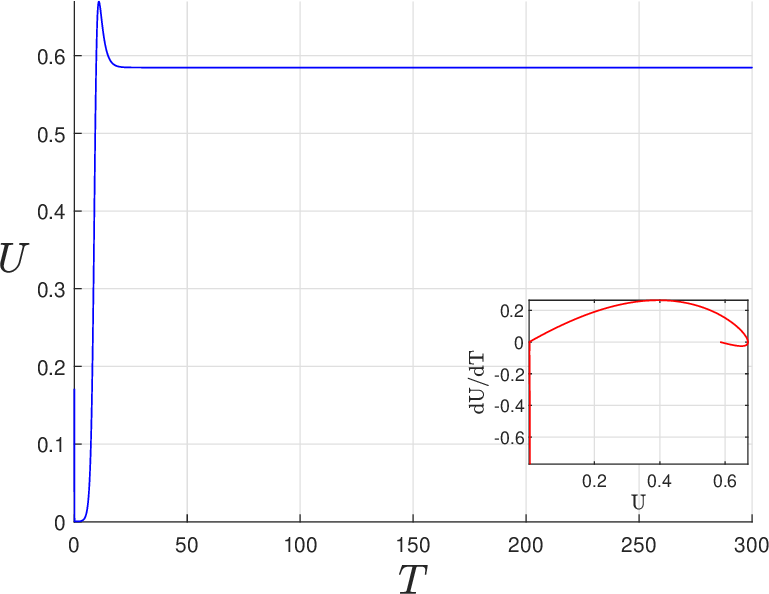} 
\includegraphics[scale=0.5]{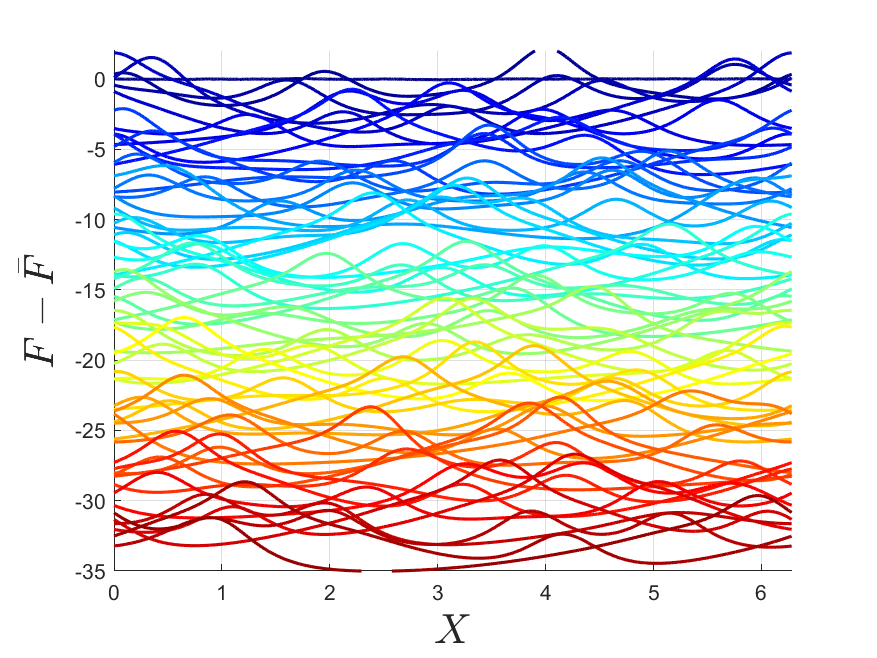} \hspace{1cm}
\includegraphics[scale=0.5]{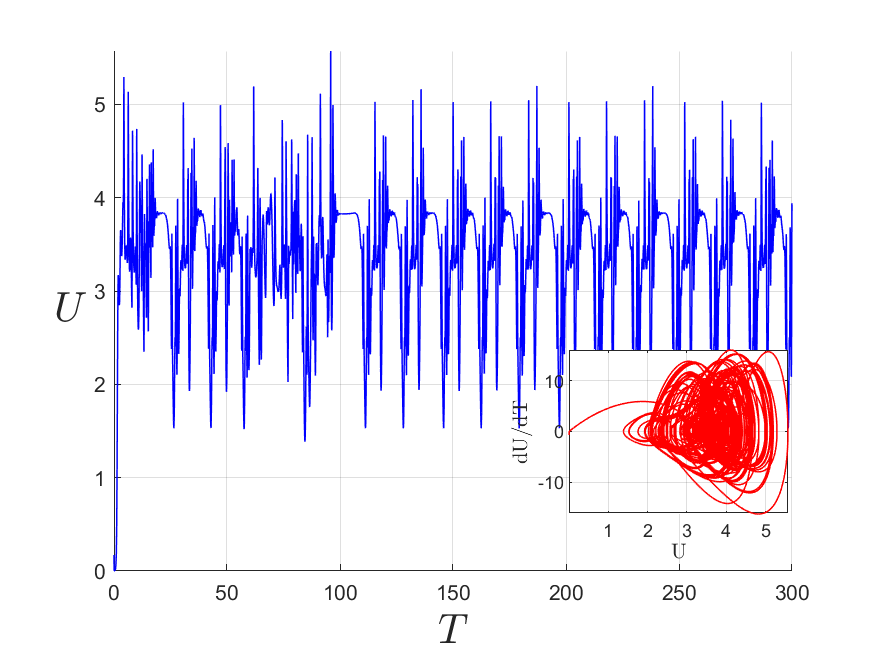} 
\caption{Numerical solution in a relatively large domain ($\nu=0.1$) with $\mu=+1$ (top row) and $\mu=-1$ (bottom row). Plotted are time snapshots of the flame shape and the trend of the propagation speed $U$; the small inset corresponds to the phase portrait, i.e., $U$ versus $dU/dT$. }
\label{fig:smalldomain}
\end{figure}

\section{Illustrative numerical results}

It is of interest to examine, by numerical means, some representative solutions of the evolution equation that has been derived. To this end, we introduce the scaling transformations
\begin{equation}
    \tau = \ep \sqrt \ep t/\sqrt{\mc N}, \qquad \xi = \sqrt \ep x/\sqrt{\mc N}, \qquad F = f/\sqrt{\ep\mc N}
\end{equation}
by which the equation~\eqref{DLDT} is reduced to the form
\begin{equation}
    F_\tau + \frac{1}{2}F_\xi^2 = \mu F_{\xi\xi} + \mathcal{H}(F_\xi) +  \mathcal{H}(F_{\xi\xi\xi}). \label{DLDT1}
\end{equation}
The domain is taken to be periodic, of length $2\pi L$; in dimensional units this corresponds to a length of order $ L\times\delta_L/\sqrt{\ep}$. From a numerical standpoint, it is convenient to to work with the slope variable $G=F_\xi$;  in this formulation the gradual drift of the mean value of $F$  is avoided~\cite{rajamanickam2024hydrodynamic}. Furthermore, the computational domain may be held fixed by introducing the variables $X=\xi/L$ and $T=\tau/L$ such that
\begin{equation}
    G_T + GG_X = \mu \sqrt\nu\, G_{XX} + \mathcal{H}(G_X) +  \nu\,\mathcal{H}(G_{XXX}), \label{Geq}
\end{equation}
where $\mu=\mc M/\sqrt\ep\sim O(1)$ provides a measure of the Markstein number and $\nu=1/L^2>0$ is inversely proportional to the square of the domain size. The initial and boundary conditions are given by
\begin{equation}
    G(X,0)=G_0(X), \qquad G(X,T)=G(X+2\pi,T). \label{bc}
\end{equation}
From the solution for $G(X,T)$, one may define a global propagation speed $U(T)$, interpretable as the gradient energy of the flame front, by
\begin{equation}
    U(T) =   \tfrac{1}{2}\overline{ F_X^2}= \tfrac{1}{2}\overline{ G^2} \quad \text{where} \quad \overline{\varphi}=\frac{1}{2\pi}\int_0^{2\pi}\varphi\, dX.
\end{equation}

In the numerical computations, two values of $\mu$ are selected. One,  $\mu=+1$, corresponds to a stabilising DT mechanism and the other, $\mu=-1$, to a destabilising one. It should be noted that although $\mu $ changes sign, the magnitude of the Markstein number varies only slightly. The initial condition for $G_0$ is taken to be a random function generated with seed 42 for reproducibility. The numerical results are discussed below using flame front evolution, accompanied by the $U(T)$ dynmaics. Readers will find it beneficial to consult the animations of the flame front evolution provided in the supplementary material, which vividly illustrate the complex dynamics described in the figures.

Figures~\ref{fig:largedomain} and~\ref{fig:smalldomain} display the solutions obtained for a relatively large domain ($\nu=0.01$) and a smaller domain ($\nu=0.1$), respectively. The contrast between the two cases is striking. When $\mu=-1$, the flame front exhibits fine cellular structures, larger propagation speeds, and a distinctly chaotic evolution. In contrast, when $\mu=+1$, the flame develops larger cells with isolated cusps, and the dynamics appears regular and ordered. These two distinct regimes are evidently analogous to the behaviour of the Michelson--Sivashinsky equation, which describes the pure Darrieus--Landau instability, and the Kuramoto--Sivashinsky equation, which characterises systems with anti-diffusion and fourth-order stabilisation. Remarkably, both are captured within the present unified framework, governed by the single crossover equation~\eqref{DLDT}. 

\begin{figure}[h!]
\centering
\includegraphics[scale=0.5]{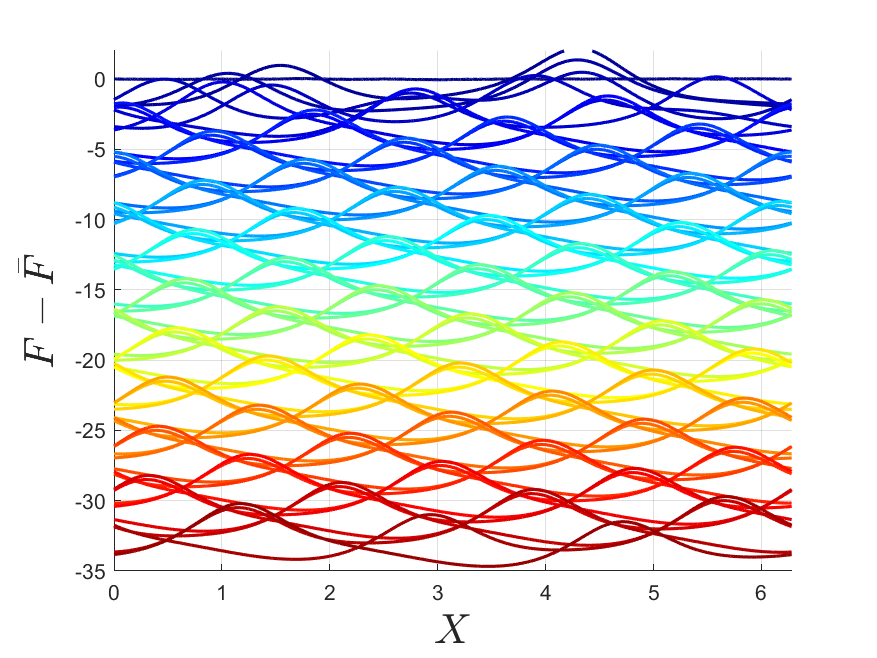} \hspace{1cm}
\includegraphics[scale=0.5]{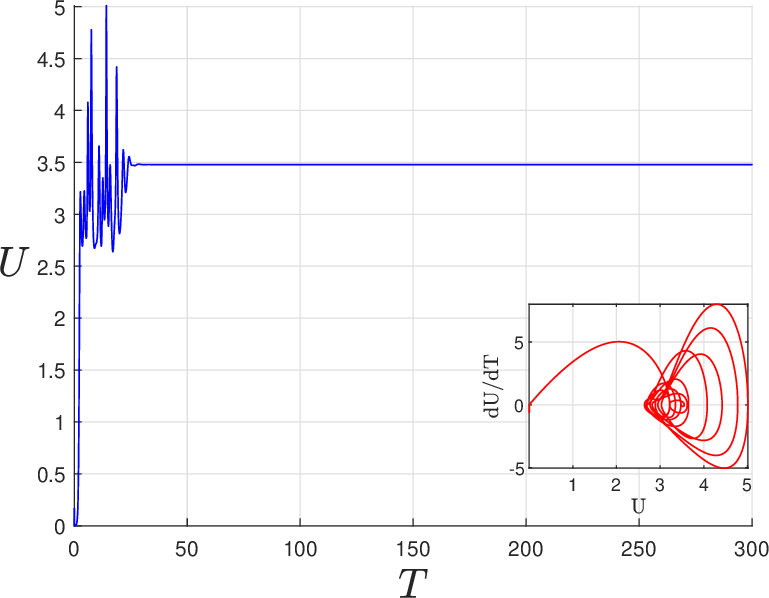} 
\includegraphics[scale=0.5]{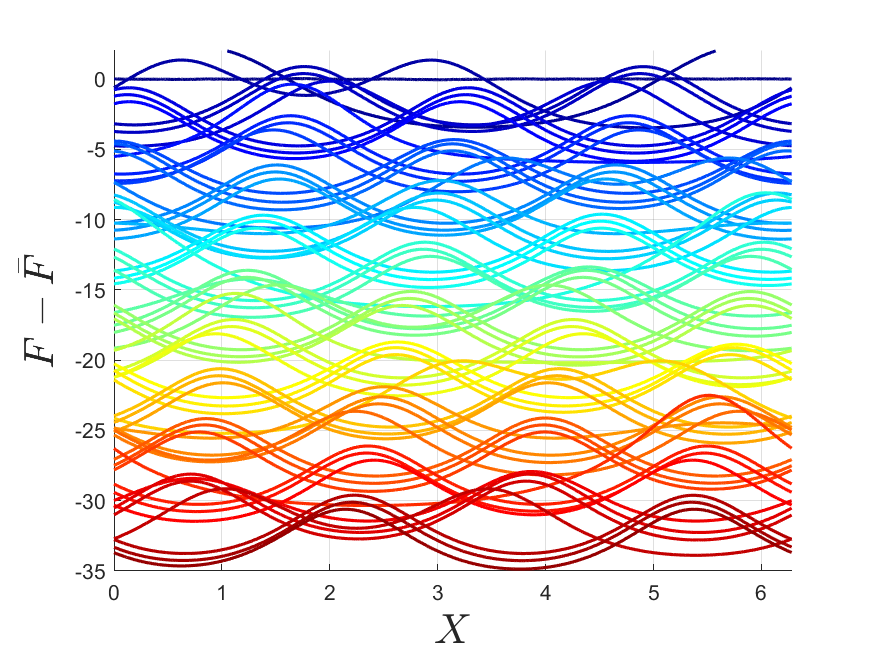} \hspace{1cm}
\includegraphics[scale=0.5]{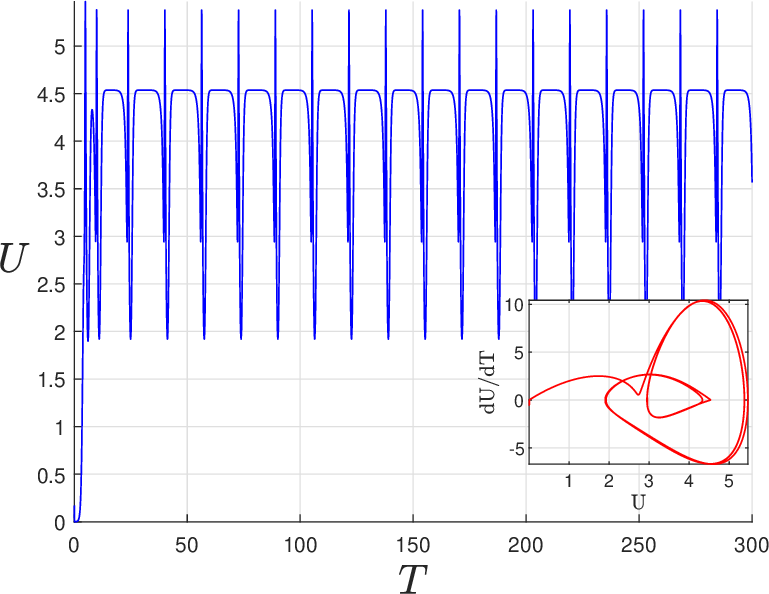} 
\includegraphics[scale=0.5]{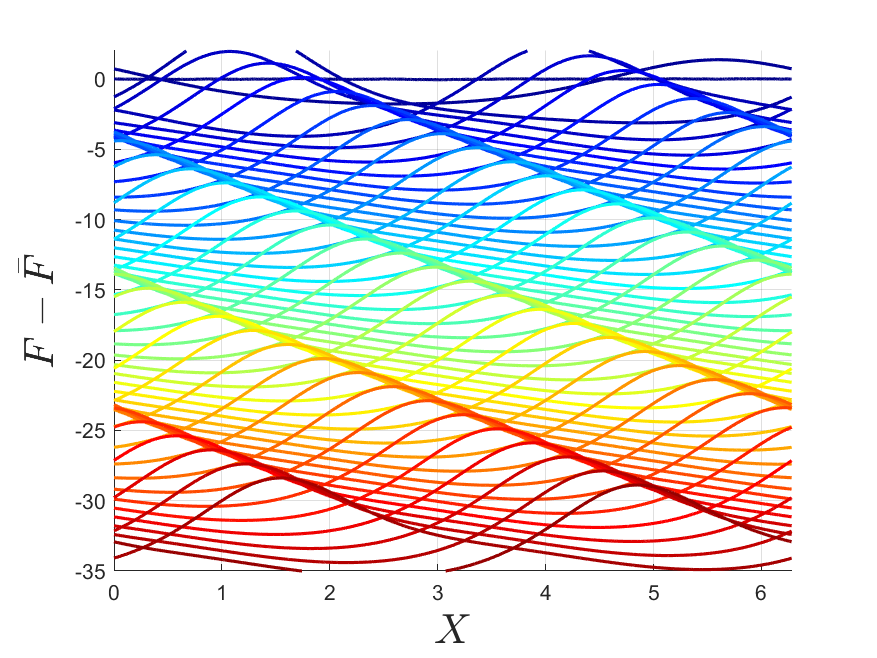} \hspace{1cm}
\includegraphics[scale=0.5]{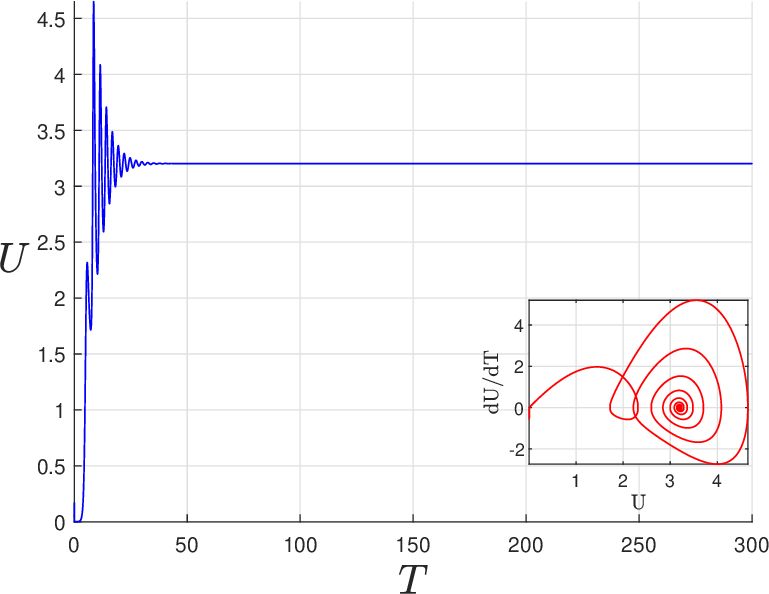} 
\caption{Numerical solution in a relatively small domains, $\nu=0.2$ (top row), $\nu=0.4$ (middle row) and $\nu=0.6$ (bottom row) with $\mu=-1$. The phase portrait for $\nu=0.4$ shows a homoclinic orbit connecting stable and unstable saddle points at $(U,dU/dT)=(4.54,0)$.}
\label{fig:more}
\end{figure}

\begin{figure}[h!]
\centering
\includegraphics[scale=0.5]{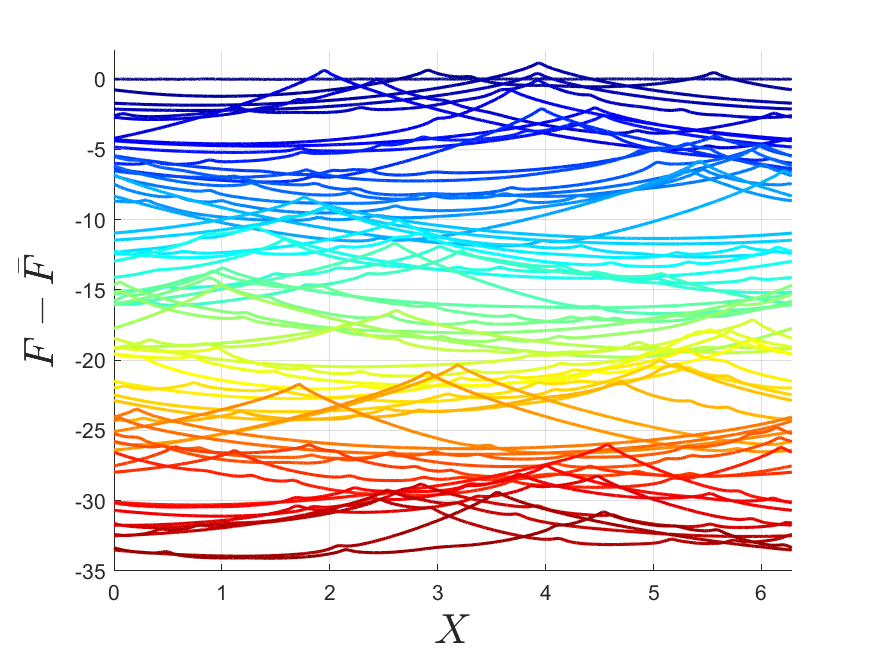} \hspace{1cm}
\includegraphics[scale=0.5]{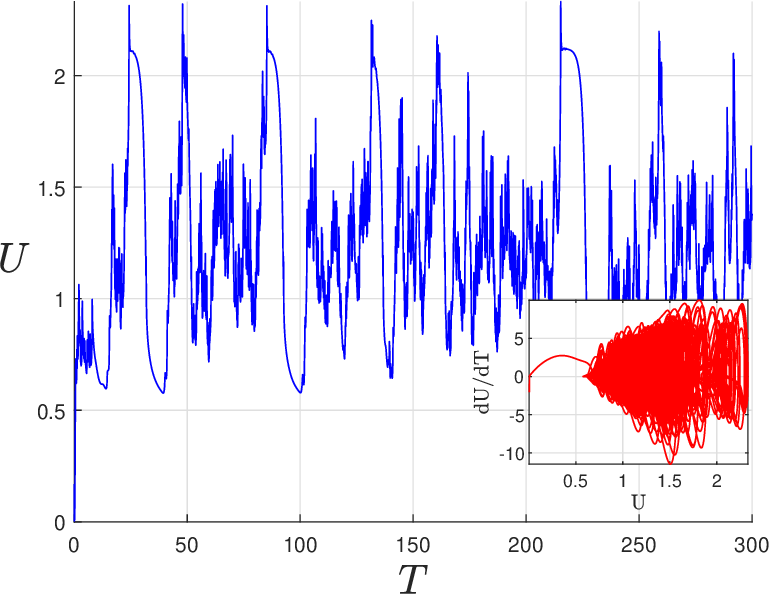} 
\caption{Numerical solution in a large domain, $\nu=0.0005$, with $\lambda=+1$. Even for positive Markstein numbers, the dynamics becomes chaotic for sufficiently large domain sizes. Remarkably, the chaotic state retains the characteristic single-cusp structure of the DL instability, but the cusp is intermittently destroyed by the emergence of short-wavelength wrinkles and then reforms, giving rise to a complex, recurrent cycle. The supplementary file containing the animation is more insightful.}
\label{fig:last}
\end{figure}

For even smaller domains, the $\mu=+1$ case does not yield further interesting dynamics, provided $\nu \gtrapprox 10^{-3}$, whereas the $\mu=-1$ case continues to exhibit rich dynamic behaviour, as illustrated for three values of $\nu\sim 10^{-2}$ in Fig.~\ref{fig:more}. To summarise, the new proposed equation~\eqref{DLDT} exhibits fundamentally different characteristics depending on the sign of the Markstein number. For  $\mc M>0$ (i.e., $\mu>0$), the dynamics resembles that of the Michelson–-Sivashinsky equation,  though it may not now be represented by a finite number of poles and cusp coalescence; this is true for moderately large domains, because in very large domains, chaos appears even when $\mc M>0$, as demonstrated in Fig.~\ref{fig:last}. The last observation may perhaps have some relevance to the fractal structures reported in~\cite{yu2015fractal} for pure DL instability. For $\mc M<0$ (i.e., $\mu<0$), the behaviour is akin to that of the Kuramoto–-Sivashinsky equation, corresponding to an effectively infinite-dimensional dynamical system with sustained chaotic activity. 

In sufficiently large domains, regardless of the sign of $\lambda$, two competing effects are constantly at play: the tendency to form large-scale DL cusp structures and the propensity to generate small-scale cellular wrinkles. These two tendencies perpetually interact, each influencing and often overtaking the other, either sequentially or simultaneously. The resulting dynamics is thus a ceaseless competition between coherent structure formation and fine-scale disruption, giving rise to the complex, recurrent cycles observed in the simulations. The present model thus provides a unified description spanning both regimes, bridging the gap between two classical paradigms of flame front instability while also revealing this richer, emergent behaviour.

\section{Can the cubic term be destabilising?}

After this brief note was written, the author was made aware of the latest publication by Bechtold and Matalon~\cite{bechtold2026long}, which similarly explores a coupled model emphasizing a cubic term. Their derivation, based on the standard multi-scale theory, reveals an unexpected sensitivity to the Prandtl number (Fig.~3 in~\cite{bechtold2026long}). Specifically, for realistic Prandtl numbers (e.g., $\Pra=0.7$), their cubic correction is destabilising for all Lewis numbers of practical interest, including at $Le=1$, indicating a new type of instability on top of DL and DT. A similar implication was hinted at in the earlier work of Class \textit{et al.}~\cite{class2003unified}, who alluded to a negative surface-tension-like effect within the same multi-scale framework. While these derivations represent valuable steps toward a unified description, it is important to recognise that the MMCJ theory is fundamentally a long-wave expansion, formally valid only for small wavenumbers, with the small parameter being the ratio of flame thickness to hydrodynamic scale. Extending this expansion to cubic terms may not be rigorously justified; the theory's limit may have already been reached at the quadratic level, and cubic-order predictions, whether stabilising or destabilising, should be interpreted with caution. Thus, the prediction of a cubic destabilisation suggests that either there is a new instability lurking, which has not been delineated from DL in experiments and simulations, or a careful re-examination of the classical multi-scale framework is required, especially in the DL-DT crossover regime. A non-standard asymptotic analysis may be needed to resolve the sign of this term.

From the phenomenological perspective adopted here, if a cubic term is taken to be destabilising, the system's saturation would necessarily rely on higher-order quartic terms, leading to the scaling laws discussed in Section~3.3 in the weak heat-release approximation. This requires a specific (codimension-two) alignment where $\mc M$ and $\mc N$ vanish simultaneously, i.e.,
\begin{equation}
    \mc N= \mc N(\mc M) \quad \text{such that} \quad \mc N(\mc M\to 0)\to 0 \quad \text{when} \quad \ep\ll 1.
\end{equation}
In the present framework, however, the cubic term is interpreted as a direct hydro-diffusive coupling that persists independently of the product $\epsilon \mathcal{M}$. By defining the coupling constant as $-(C_0 + C_1 \epsilon \mathcal{M})|k|^3$, the present model targets the regime where $C_0 \neq 0$ is dominant, thereby providing a robust stabilising mechanism. This stands in contrast to the case where $C_0 \ll 1$, in which both the cubic  and quadratic terms vanish together, requiring a quartic term ($-4k^4$) for stabilisation. The present work advocates for $C_0 \neq 0$ as the physically immediate crossover regime, representing a strong, non-perturbative coupling between hydrodynamics and the internal flame structure. This interaction is potentially invisible to standard perturbative asymptotics, which treat the ratio of flame thickness to hydrodynamic length as a small parameter.  The possibility $C_0 \ll 1$, and the associated deep asymptotic limit of Section 3.3, is thus left as a restricted, fine-tuned (codimension-two) special case.

\section{Concluding remarks}

A minimal phenomenological model coupling Darrieus--Landau and diffusive-thermal instabilities in premixed flames has been proposed. Starting from a modified dispersion relation that captures their leading-order coupling, two asymptotically distinct regimes are identified. The classical Michelson--Sivashinsky equation emerges when the Markstein number is order unity and positive, where flames exhibit order-unity amplitudes and long wavelengths. In the distinguished crossover regime where both instabilities interact at equal order and Markstein number is small, a generalized Michelson--Sivashinsky equation with an additional nonlocal term is obtained. This regime produces finer cellular structures, faster growth rates, and smaller amplitudes — a direct signature of coupled DL-DT dynamics. For the strongly unstable case where the Markstein number is negative, a heuristic model bridging the Michelson--Sivashinsky and Kuramoto--Sivashinsky equations is proposed via a short-wavelength cutoff, acknowledging its phenomenological nature. This simple framework captures essential physics and offers a tractable starting point for numerical exploration or extensions incorporating gravity, heat losses, or confinement. 

\begin{figure}[h!]
\centering
\includegraphics[scale=0.55]{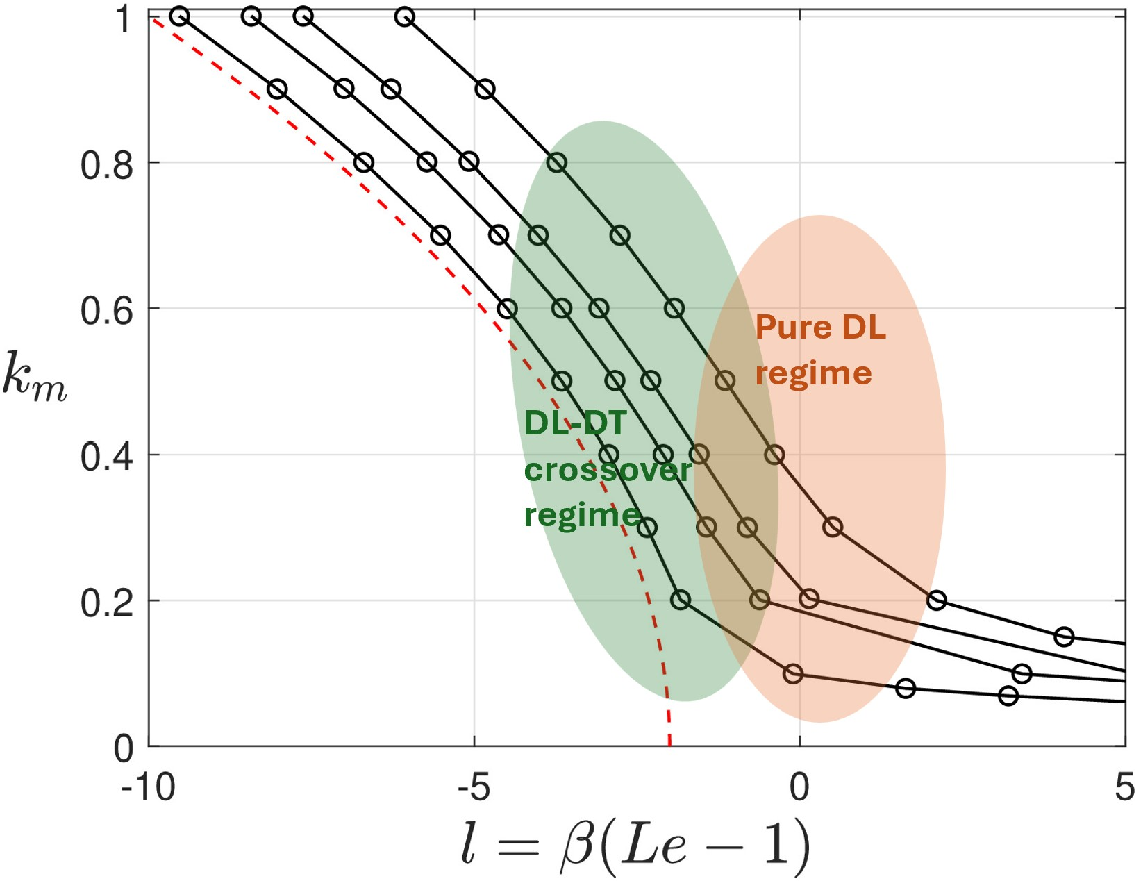} 
\caption{Marginal stability curves for different values of $\epsilon=(r-1)/2$ in the limit $\beta\to \infty$ where $\beta$ is the Zeldovich number. The red dashed line corresponds to the formula $l=-2-8k^2$, applicable when $r=1$ (i.e., $\epsilon=0$), obtained by Sivashinsky~\cite{sivashinsky1977diffusional}. The four solid lines correspond to $r=\{1.2,1.5,2,5\}$ (i.e., $\epsilon=\{0.1,0.25,0.5,2.0\}$), obtained by Jackson and Kapila~\cite{jackson1984effect}. The green shaded region indicates the DL--DT crossover regime where the cubic term $-\mathcal{A}|k|^3$ identified in the present work becomes significant; the red shaded region indicates where it is negligible.}
\label{fig:regime}
\end{figure}

A key contribution of this work is the identification of the crossover coupling term, $-\mc N|k|^3$, where $\mathcal{N} = \mc A/\delta_L^2$ is the hydro-diffusive number and $\mc A$ the corresponding hydro-diffusive area, the characteristic area over which hydrodynamic and diffusive transport processes interact.  This cubic stabilisation finds a mathematical analogue in the well-known effect of surface tension at immiscible fluid interfaces. The formal scaling $\mc M \sim \sqrt{\ep\mc N}$ delineates a distinguished regime in which the two instabilities participate on an equal footing. Although the expansion parameter $\ep$ is not asymptotically small under conditions commonly encountered in experimental flames, this scaling nevertheless provides a useful theoretical guide for identifying when coupled effects become significant. To render the physical content of this coupling more transparent, we may write the dispersion relation for the DL-DT crossover in its dimensional form (see Figure~\ref{fig:regime}): 
\begin{equation}
    \frac{\sigma}{S_L} = \frac{1}{2}\left(\frac{\rho_u}{\rho_b}-1\right)  |k| - \mc L k^2 - \mc A|k|^3, \qquad \text{with} \qquad \mc L \sim \sqrt{\left(\frac{\rho_u}{\rho_b}-1\right) \frac{\mc A}{2}}.
\end{equation}
In this expression, $\mc A=\mc N\delta_L^2$ appears explicitly as an area, clarifying its role as the spatial scale of hydro-diffusive interaction, while $\mc L= \mc M\delta_L$ retains its familiar interpretation as the Markstein length. This cubic term, absent in classical treatments, emerges naturally as the leading-order manifestation of the interaction between the long-wave Darrieus--Landau instability, $\ep|k|$, and and the short-wave diffusive-thermal instability, $-\mc M k^2$. There exists no physical or mathematical justification for supposing the coefficient $\mc N$ to be zero; rather, its omission in the past is a consequence of treating these instabilities in isolation. By moving beyond the single-parameter asymptotic limit, wherein either $\epsilon$ or $\mc M$ is presumed small, we have demonstrated that this hydro-diffusive term provides a critical nonlocal stabilisation that persists even in the limit of a vanishing Markstein number.  The evolution equation in general form can be written as
\begin{equation}
    f_t +\tfrac{1}{2}(\nabla f)^2 = \left[(-\Delta)^{1/2} -\mu(-\Delta)^1 -(-\Delta)^{3/2}\right] f, \qquad \mu \in \mathbb R.
\end{equation}
It is perhaps to be expected that the familiar pole decomposition, which so elegantly characterises the Michelson--Sivashinsky equation,  does not extend to this equation. The $\tfrac{3}{2}$-Laplacian, $ -(-\Delta)^{3/2}f$, fundamentally alters the singularity structure of the solutions, precluding a finite-pole representation. This suggests that the dynamics encompassed by the present equation may be considerably richer than those of the pure DL case, in which the asymptotic state is typically governed by the coalescence of poles into a single cusp. This mathematical complexity mirrors the physical intricacy of the regime under consideration. By incorporating the coupling between hydrodynamic and diffusive-thermal effects, the proposed model offers a more robust framework for interpreting the fine-scale cellular structures that have been observed in experiments on flames near the threshold of stability.  Within this distinguished crossover limit, the hydro-diffusive area $\mc A$  serves as a robust coupling constant. Whereas the Markstein number $\mc M$  is highly sensitive to the Lewis number, changing sign as the DT instability threshold is crossed, the coefficient $\mc N$ remains a stable, leading-order interaction term. To a first approximation, $\mc N$ may be treated as independent of the Lewis number in this narrow regime; however, it may be expected to vary as the internal flame structure undergoes more substantial modifications farther from the threshold.  A first-principles derivation of the coefficient $\mathcal{N}$ from the full Navier--Stokes equations remains a challenging undertaking, requiring an asymptotic development that goes beyond the established Matalon--Matkowsky--Clavin--Joulin theory. Its determination may prove more tractable within the framework of Darcy's law, where the hydrodynamic description is considerably simplified~\cite{rajamanickam2024hydrodynamic,rajamanickam2026flame}.  Alternatively, $\mathcal{N}$ might be inferred from experimental measurements of cellular flame structures or extracted from direct numerical simulations, thereby providing a practical route to quantifying the hydro-diffusive area $\mc A$. More specifically, the limit $\mc M\to 0^\pm$ is of particular interest; in this limit, the classical theory, with its assumption of an order-unity Markstein number, becomes inapplicable. A new hydrodynamic description of premixed flames is therefore required, a direction that is both warranted and reserved for future investigation.

Note that the present study has been conducted within the framework of a weak-heat-release approximation, adopted in the interest of simplicity. The qualitative features of the discussion may be expected to carry over, with appropriate modifications, when this approximation is relaxed. In the latter case, for instance, the quadratic stabilisation vanishes not simply when $\mc M \leq 0$, but rather when $\mc M-\mc M_c \leq 0$, where $\mc M_c$ denotes a critical Markstein number influenced by effects such as viscosity (see, for example, formula (19)  and $\S$3.2 in~\cite{matalon2018darrieus}, which follows from a Taylor-series expansion of the Clavin--Garcia equation).

\begin{figure}[h!]
\centering
\includegraphics[scale=0.55]{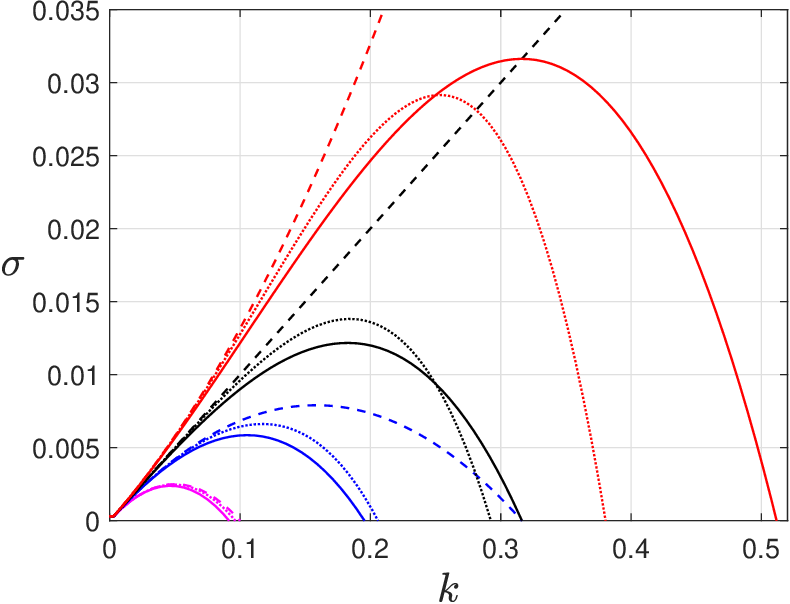} 
\caption{Comparison of dispersion curves for the Sivashinsky model (dotted), the present model (solid), and the pure DL instability (dashed) for $\ep=0.1$, $\mc N=1$ and four values of the Markstein number: $\mc M=1$ (magenta), $\mc M=\sqrt\ep$ (blue), $\mc M=0$ (black) and $\mc M=-\sqrt\ep$ (red).}
\label{fig:disp}
\end{figure}

Finally, it is appropriate to conclude this note by comparing (see Fig.~\ref{fig:disp}) the distinguished DL-DT crossover regime proposed by Sivashinsky~\cite{sivashinsky1988nonlinear,matalon2007intrinsic} for small values of $\mc M$ (or $\mc M-\mc M_c$) with ours:
\begin{align}
    \text{Sivashinsky:} &\quad  \sigma = \ep |k| - \mc M k^2 - 4k^4 \quad &&\Rightarrow \quad k \sim \ep^{1/3}, \quad \sigma \sim \ep ^{4/3}, \quad \left\{f,\mc M\right\} \sim \ep^{2/3}, \nonumber \\   
    \text{Current model:} &\quad  \sigma = \ep |k| - \mc M k^2 - \mc N |k|^3, \quad &&\Rightarrow \quad k\sim \sqrt\ep, \quad \sigma \sim \ep\sqrt\ep, \qquad \left\{f,\mc M\right\}\sim \sqrt\ep, \nonumber 
\end{align}
or, alternatively, 
\begin{align}
    \text{Sivashinsky:} &\quad  \sigma = \ep |k| - \mc M k^2 - 4k^4 \quad &&\Rightarrow \quad k \sim \sqrt{\mc M}, \quad \sigma \sim \mc M^2, \quad f\sim \mc M, \quad \ep \sim \mc M^{3/2}, \nonumber \\   
    \text{Current model:} &\quad  \sigma = \ep |k| - \mc M k^2 - \mc N |k|^3, \quad &&\Rightarrow \quad k \sim \mc M, \quad \sigma \sim \mc M^3, \quad f\sim \mc M, \quad \ep \sim \mc M^2. \nonumber 
\end{align}
These scalings complement the pure DL regime scalings:
\begin{equation}
     \text{Pure DL:} \quad  \sigma = \ep |k| - \mc M k^2, \quad \Rightarrow \quad k \sim \ep, \quad \sigma \sim \ep^2, \quad \left\{f,\mc M\right\} \sim 1. \nonumber
\end{equation}
The reader may also refer to a recent study~\cite{creta2020propagation} that explored nonlinear flame dynamics based on Sivashinsky's scalings. Since our Markstein-number scaling $\mc M \sim \sqrt\ep$ characteristic of the present model exceeds that of Sivashinsky $\mc M\sim \ep^{2/3}$, the cubic stabilisation becomes operative earlier as the Markstein number is reduced toward the DT instability threshold. In other words, as a flame transitions from DL-modulated propagation toward the DT instability threshold, the cubic term engages first, providing the primary stabilising mechanism, while the quartic term only emerges in the deeper asymptotic limit where $\mc M$ is extremely small. Accordingly, when comparing cubic $(-|k|^3)$ versus quartic $(-k^4)$ stabilisation, the cubic model yields cellular structures of larger characteristic dimension, slower growth rates, and greater amplitudes.  The present scaling thus captures the physically more immediate crossover regime where both hydrodynamic and diffusive-thermal instabilities participate on an equal footing.



\bibliographystyle{elsarticle-num}
\bibliography{elsarticle-template}

@article{landau1944slow,
  title={On the theory of slow combustion},
  author={Landau, L D},
  journal={Zh. Eksp. Teor. Fiz},
  volume={14},
  number={6},
  pages={240--244},
  year={1944}
}

@article{darrieus1938propagation,
  title={Propagation d’un front de flamme},
  author={Darrieus, G},
  journal={La Technique Moderne},
  volume={30},
  number={18},
  year={1938},
  publisher={Le Congr{\`e}s de M{\'e}canique Appliqu{\'e}e}
}

@article{matalon2007intrinsic,
  title={Intrinsic flame instabilities in premixed and nonpremixed combustion},
  author={Matalon, Moshe},
  journal={Annu. Rev. Fluid Mech.},
  volume={39},
  number={1},
  pages={163--191},
  year={2007},
  publisher={Annual Reviews}
}

@book{zeldovich1944theory,
  title={Theory of combustion and detonation of gases},
  author={Zeldovich, Ya. B},
  year={1944},
  publisher={Headquarters, Wright-Patterson Air Force Base, Air Materiel Command}
}

@article{clavin1983premixed,
  title={Premixed flames in large scale and high intensity turbulent flow},
  author={Clavin, P and Joulin, G},
  journal={J. Phys. (Paris), Lett.},
  volume={44},
  number={1},
  pages={1--12},
  year={1983},
  publisher={Les Editions de Physique}
}

@article{matalon1982flames,
  title={Flames as gasdynamic discontinuities},
  author={Matalon, Moshe and Matkowsky, Bernard J},
  journal={J. Fluid Mech.},
  volume={124},
  pages={239--259},
  year={1982},
  publisher={Cambridge University Press}
}

@book{clavin2016combustion,
  title={Combustion waves and fronts in flows: flames, shocks, detonations, ablation fronts and explosion of stars},
  author={Clavin, Paul and Searby, Geoff},
  year={2016},
  publisher={Cambridge University Press}
}

@book{williams2018combustion,
  title={Combustion theory},
  author={Williams, Forman A},
  year={1985},
  publisher={CRC Press}
}

@article{matalon2018darrieus,
  title={The {D}arrieus--{L}andau instability of premixed flames},
  author={Matalon, Moshe},
  journal={Fluid Dynamics Research},
  volume={50},
  number={5},
  pages={051412},
  year={2018},
  publisher={IOP Publishing}
}

@article{clavin1982effects,
  title={Effects of molecular diffusion and of thermal expansion on the structure and dynamics of premixed flames in turbulent flows of large scale and low intensity},
  author={Clavin, Paul and Williams, F A},
  journal={J. Fluid Mech.},
  volume={116},
  pages={251--282},
  year={1982},
  publisher={Cambridge University Press}
}

@incollection{pelce1988influence,
  title={Influence of hydrodynamics and diffusion upon the stability limits of laminar premixed flames},
  author={Pelce, Pierre and Clavin, Paul},
  booktitle={Dynamics of curved fronts},
  pages={425--443},
  year={1988},
  publisher={Elsevier}
}

@article{clavin1985effect,
  title={Effect of heat losses on the limits of stability of premixed flames propagating downwards},
  author={Clavin, P. and Nicoli, C.},
  journal={Combust. Flame},
  volume={60},
  number={1},
  pages={1--14},
  year={1985},
  publisher={Elsevier}
}

@article{sivashinsky1988nonlinear,
  title={Nonlinear analysis of hydrodynamic instability in laminar flames—{I}. {D}erivation of basic equations},
  author={Sivashinsky, G I},
  journal={Acta Astronaut.},
  volume={4},
  number={11-12},
  pages={1177--1206},
  year={1977},
  publisher={Elsevier}
}

@article{rajamanickam2024tricritical,
  title={Tricritical point as a crossover between type-{I}s and type-{II}s bifurcations},
  author={Rajamanickam, Prabakaran and Daou, Joel},
  journal={Prog. Scale Model. Int. J.},
  volume={4},
  pages={1--7},
  year={2023},
  publisher={UK}
}

@article{rajamanickam2026flame,
  title={Flame dynamics and {M}arkstein numbers in {H}ele-{S}haw cells and porous media under Darcy's law},
  author={Rajamanickam, Prabakaran and Daou, Joel},
  year={2026}
}

@article{rajamanickam2024hydrodynamic,
  title={Hydrodynamic theory of premixed flames under {D}arcy's law},
  author={Rajamanickam, Prabakaran and Daou, Joel},
  journal={Phys. Fluids},
  volume={36},
  number={12},
  pages={1--7},
  year={2024},
  publisher={AIP Publishing}
}

@article{denet1995numerical,
  title={A numerical study of premixed flames {D}arrieus-{L}andau instability},
  author={Denet, B and Haldenwang, P},
  journal={Combust. Sci. Technol.},
  volume={104},
  number={1-3},
  pages={143--167},
  year={1995},
  publisher={Taylor \& Francis}
}

@article{yu2015fractal,
  title={Fractal flame structure due to the hydrodynamic {D}arrieus--{L}andau instability},
  author={Yu, Rixin and Bai, Xue-Song and Bychkov, Vitaly},
  journal={Phys. Rev. E},
  volume={92},
  number={6},
  pages={063028},
  year={2015},
  publisher={APS}
}

@article{bechtold2026long,
  title={A long-wave approximation of the dispersion relation encompassing hydrodynamic and diffusional-thermal instabilities for premixed flames},
  author={Bechtold, John K and Matalon, Moshe},
  journal={Combust. Flame},
  volume={288},
  pages={114951},
  year={2026},
  publisher={Elsevier}
}

@article{class2003unified,
  title={A unified model of flames as gasdynamic discontinuities},
  author={Class, Andreas G and Matkowsky, B J and Klimenko, A Y},
  journal={J. Fluid Mech.},
  volume={491},
  pages={11--49},
  year={2003},
  publisher={Cambridge University Press}
}

@article{creta2020propagation,
  title={Propagation of premixed flames in the presence of {D}arrieus--{L}andau and thermal diffusive instabilities},
  author={Creta, Francesco and Lapenna, Pasquale Eduardo and Lamioni, Rachele and Fogla, Navin and Matalon, Moshe},
  journal={Combust. Flame},
  volume={216},
  pages={256--270},
  year={2020},
  publisher={Elsevier}
}

@article{denet1992numerical,
  title={Numerical study of thermal-diffusive instability of premixed flames},
  author={Denet, Bruno and Haldenwang, Pierre},
  journal={Combust. Sci. Technol.},
  volume={86},
  number={1-6},
  pages={199--221},
  year={1992},
  publisher={Taylor \& Francis}
}

@article{daou2024diffusive,
  title={Diffusive-thermal instabilities of a planar premixed flame aligned with a shear flow},
  author={Daou, Joel and Rajamanickam, Prabakaran},
  journal={Combust. Theory Model.},
  volume={28},
  number={1},
  pages={20--35},
  year={2024},
  publisher={Taylor \& Francis}
}

@article{jackson1984effect,
  title={Effect of thermal expansion on the stability of plane, freely propagating flames},
  author={Jackson, T L and Kapila, A K},
  journal={Combust. Sci. Technol.},
  volume={41},
  number={3-4},
  pages={191--201},
  year={1984},
  publisher={Taylor \& Francis}
}

@article{sivashinsky1977diffusional,
  title={Diffusional-thermal theory of cellular flames},
  author={Sivashinsky, G I},
  journal={Combust. Sci. Technol.},
  volume={15},
  number={3-4},
  pages={137--145},
  year={1977},
  publisher={Taylor \& Francis}
}

\end{document}